\documentclass[prb,twocolumn,showpacs,aps,superscriptaddress,floatfix]{revtex4}
\usepackage{amsmath}
\usepackage{amssymb}
\usepackage{amsfonts}
\usepackage{bm}
\usepackage{graphicx}
\usepackage{color}
\usepackage[svgnames]{xcolor}
\usepackage{soul}
\usepackage{hyperref}
\usepackage{verbatim}

\DeclareMathOperator{\arccosh}{arccosh}
\DeclareMathOperator{\diag}{diag}
\DeclareMathOperator{\sign}{sign}

\begin{document}

\title{Competition of spatially inhomogeneous states in antiferromagnetic
Hubbard model}

\author{S.\,V.~Kokanova}
\affiliation{Skolkovo Institute of Science and Technology, Moscow}

\author{P.\,A.~Maksimov}
\affiliation{Bogolyubov Laboratory of Theoretical Physics,
Joint Institute for Nuclear Research, Dubna, Moscow region 141980, Russia}

\author{A.\,V.~Rozhkov}
\affiliation{Institute for Theoretical and Applied Electrodynamics, Russian
Academy of Sciences, 125412 Moscow, Russia}

\author{A.\,O.~Sboychakov}
\affiliation{Institute for Theoretical and Applied Electrodynamics, Russian
Academy of Sciences, 125412 Moscow, Russia}

\begin{abstract}
In this work we study zero-temperature phases of the anisotropic Hubbard
model on a three-dimensional cubic lattice in a weak coupling regime. It is
known that, at half-filling, the ground state of this model is
antiferromagnetic (commensurate spin-density wave). For non-zero doping,
various types of spatially inhomogeneous phases, such as phase-separated
states and the state with domain walls (``soliton lattice"), can emerge.
Using the mean-field theory, we evaluate the free energies of these phases
to determine which of them could become the true ground state in the limit
of small doping. Our study demonstrates that the free energies of all
discussed states are very close to each other. Smallness of these energy
differences suggests that, for a real material, numerous factors,
unaccounted by the model, may arbitrary shift the relative stability of the
competing phases. This further implies that purely theoretical prediction
of the true ground state in a particular many-fermion system is unreliable.
\end{abstract}



\date{\today}

\maketitle
\section{Introduction}

\subsection{Motivation}

Investigations of inhomogeneous electronic states of different types
constitute an active research area of modern condensed matter
physics~\cite{zheng2000observation,tranquada1995evidence,
bianconi1996determination,li2019pnictides_exp,dho2003strain,
miao2020direct,zhu2016chemical, iwaya2011stripe,sing2017influence,
laplace2012nanoscale,kokanova2019disorder,luo_dagotto2014hubb,
Bianconi2015phaSep, sboychakov2013dmft_hub-I, ourPRB_phasepAFM2017,
Igoshev2015hubbard,sheehy_cold_atoms2007,narayanan2014coexistence}.
In this work, we study inhomogeneous states arising in the weakly
doped antiferromagnetic (AFM) insulator. This physical situation is of
interest in the context of
cuprates~\cite{kivelson2003detect},
pnictides~\cite{park2009electronic},
as well as other antiferromagnetic
systems~\cite{narayanan2014coexistence}.
(Some non-magnetic superconducting systems may
demonstrate~\cite{sheehy_cold_atoms2007}
similar instabilities as well.)
For the doped AFM materials inhomogeneities of different types, such as
``stripes" (domain walls), phase separation, and ``checkerboard" state, are
discussed (see, e.g.,
Refs.~\onlinecite{tokatly1992,kivelson2003detect,
park2009electronic,igoshev2010,laplace2012nanoscale,
dolgirev_checkerboard2017,checkerboard_fine2019,Fine2011checkerboard,
narayanan2014coexistence,pressure_phasep_rakhmanov2020}).

Theoretically, many-body states hosting these types of inhomogeneities are
quite common, and can be studied with the help of standard
approximations~\cite{buzdin_tugushev_soliton_latt1983,
ourPRB_phasepAFM2017, sboychakov2013dmft_hub-I, igoshev2010,
Igoshev2015hubbard}.
Often, various inhomogeneous phases compete against each other to become
the true ground state of a model Hamiltonian. The outcome of this
competition is usually
presented~\cite{buzdin_tugushev_soliton_latt1983,igoshev2010, ourPRB_phasepAFM2017,Igoshev2015hubbard}
as a phase diagram that depicts how various states, both homogeneous and
inhomogeneous, replace each other upon parameters variations. Yet, when
comparing the model phase diagram against experimental data, one
unavoidably has to address the following question: to which extent the
diagram calculated within a simplified theoretical framework with the help
of approximate approaches is robust in the presence of various
perturbations unaccounted by the model or distorted by the approximations?

\subsection{Our results}

Answering the question formulated in the previous paragraph, it is often
implicitly assumed that, although some variability of the phase boundaries
is indeed unavoidable, the qualitative features of the diagram survive the
reality check. In our study, we examine the latter assumption and
demonstrate that the perceived stability is not automatically guaranteed at
all. Namely, it will be shown that theoretically estimated energies of
competing inhomogeneous states may be very close to each other.
Consequently, which of these phases becomes the true ground state in a real
system may be ultimately decided by factors external to the theoretical
model, such disorder, longer-range interaction, lattice effects, etc.

Specific details of our investigation are as follows. We adopt the Hubbard
model on the anisotropic cubic lattice as a study case. This model is
widely used in the theoretical literature to describe large variety of
many-fermion systems. To avoid
non-controllable approximations, we limit ourselves to the regime of weak
interactions. Stable antiferromagnetism in a weak coupling model is
possible if the Fermi surface demonstrates pronounced nesting. Indeed, it is
known that in the presence of nesting electronic liquid becomes unstable,
and an AFM phase appears at arbitrary weak interaction. For the Hubbard
model, the Fermi surface demonstrates perfect nesting at half-filling (one
electron per a lattice site) and vanishing longer-range hopping amplitudes.

In the regime of weak coupling, Hamiltonians with
nesting~\cite{rice1970,ourPRB_phasepAFM2017,ourPRL_half-met2017,
Chuang1509,PhysRevB.86.020507,PhysRevB.82.020510,vsimkovic2016ground,
mosoyan_aa_graphit2018,ourPRB_half-met2018,Nandkishore2012,
our_PRB_magfield_imp_nest,gonzalez_kohn-lutt_twist2019,
twist_graph_bias_gap_prl2018,nesting_review2017,
aa_graph2014,aa_graph2013,phasep_pnics2013,q1d2009,q1d2009_2,q1d2003,
hirschfeld_review2011,fernandez_pnic_so5_2010,gruner_book,
khokhlov2020dynamical}
allow for consistent mean-field treatment of the AFM order.
For example,
Rice~\cite{rice1970}
used such a model to study AFM state in Cr and its alloys. It was later
noted that in this class of theories, the doped AFM state is unstable with
respect to phase separation of the injected
electrons~\cite{ourPRB_phasepAFM2017}.
When the possibility of inhomogeneous state formation was taken into account,
two scenarios of phase separation were identified:
(i)~separation into the paramagnetic and commensurate AFM states and
(ii)~separation into commensurate and incommensurate AFM states.
Besides these,
(iii)~a phase with domain walls (sometimes
called~\cite{buzdin_tugushev_soliton_latt1983}
``soliton lattice") is also discussed in the
literature~\cite{schulz1989stripes,buzdin_tugushev_soliton_latt1983,
zaanen1989stripes}.

In our work, we investigate the relative stability of inhomogeneous
states~(i-iii). To determine which of them is energetically favorable, it
is necessary to compare the free energies $F$ of these states. The free
energies are calculated with the help of the mean-field approximation. As
we have mentioned above, for all three states, the values of $F$ are
virtually identical, at least at small doping. This finding and its
implications are the main focus of this paper.

Our presentation is organized as follows. In
Sec.~\ref{sec::Model}
we discuss the Hubbard model and the mean-field approach used in
calculations. In
Sec.~\ref{sec::Inhomogeneous}
we present calculations for the phase-separated states. The application of
the mean-field approximation to the state with domain walls is explained in
Sec.~\ref{sec::Domain_wall}.
Section~\ref{sec::Discussion}
is dedicated to the discussion of the results. Some auxiliary results are
relegated to two Appendices.

\section{Mean-field approach to the Hubbard model}
\label{sec::Model}

We consider antiferromagnetic state of anisotropic Hubbard model on a
three-dimensional (3D) cubic lattice in the weak coupling regime. The
Hamiltonian of the model equals to
\begin{eqnarray}
H&=&\sum_{\langle ij\rangle\sigma} t_{ij} \left(c_{i\sigma}^{\dag}c^{\phantom{\dag}}_{j\sigma} + H.c.\right)
-\mu\sum_{i\sigma}c_{i\sigma}^{\dag}c^{\phantom{\dag}}_{i\sigma}+\nonumber\\
&&+U\sum_{i}\left(n_{i\uparrow}-\frac12\right)\!\left(n_{i\downarrow}-\frac12\right),
\label{eq::Hamiltonian}
\end{eqnarray}
where
$c_{i\sigma}^{\dag}$
and
$c^{\phantom{\dag}}_{i\sigma}$
are the creation and annihilation operators for an electron with spin projection
$\sigma=\uparrow,\,\downarrow$
located in the site $i$, local density operator is
$n_{i\sigma}=c_{i\sigma}^{\dag}c^{\phantom{\dag}}_{i\sigma}$,
notation
$\langle ij\rangle$
implies that sites $i$ and $j$ are nearest neighbors, and
$t_{ij}$
represents the hopping amplitude connecting sites $i$ and $j$. Many
antiferromagnetic materials (pnictides, cuprates, organic Bechgaard salts)
demonstrate pronounced anisotropy. To model this feature, we explicitly
assume that
$t_{ij}$
are different for different orientations of the
$\langle i j \rangle$-bond:
when the bond is parallel to the
$\alpha$-axis
($\alpha=x,\,y,\,z$),
the amplitude is
$t_{ij}=t_\alpha$.

In the second term of
Eq.~\eqref{eq::Hamiltonian},
$\mu$ is the chemical potential. The last term in
Eq.~\eqref{eq::Hamiltonian}
describes on-site Coulomb repulsion of electrons with opposite spin
projections, with the interaction constant
$U > 0$. The terms $1/2$ in parentheses are added in order to chemical potential $\mu$ would equal to zero at half-filling (one electron per site).

We consider the Hubbard
model~\eqref{eq::Hamiltonian}
near the half-filling. At half-filling, the ground
state of the Hubbard model is known to be
antiferromagnetic~\cite{hubbard1995collection}.
It is assumed that small doping modifies but does not destroy the
antiferromagnetic ordering.

In the antiferromagnetic state, the averaged number of spin-up electrons on
site $i$ is not equal to the averaged number of spin-down electrons on the
same site:
$\langle n_{i\uparrow}\rangle \ne \langle n_{i\downarrow}\rangle$.
We define the position-dependent order parameter
\begin{equation}
\label{eq::mean1}
\Delta_{i}=\frac{U}{2}\left(\langle n_{i\uparrow}\rangle-\langle n_{i\downarrow}\rangle\right).
\end{equation}
For the antiferromagnetic states with domain walls, the sum
$\langle n_{i\uparrow}\rangle+\langle n_{i\downarrow}\rangle$
is also position dependent. Thus, it is useful to introduce the local doping level
\begin{equation}
\label{eq::mean2}
x_{i}=\langle n_{i\uparrow}\rangle+\langle n_{i\downarrow}\rangle-1\,.
\end{equation}
We study antiferromagnetic states of the
model~\eqref{eq::Hamiltonian}
in the weak coupling regime, when
$U<W$,
where $W$ is the bandwidth,
$W=4\sum_{\alpha}t_{\alpha}$.
In this case, the mean-field approach is the appropriate method to study
the
model~\eqref{eq::Hamiltonian}.
The mean-field scheme is based on the following replacement
\begin{equation}
\label{eq::decoupling}
n_{i\uparrow}n_{i\downarrow}\to\langle n_{i\uparrow}\rangle\,n_{i\downarrow}+\langle n_{i\downarrow}\rangle\,n_{i\uparrow}-
\langle n_{i\uparrow}\rangle\langle n_{i\downarrow}\rangle\,.
\end{equation}
Applying this substitution rule to
Hamiltonian~\eqref{eq::Hamiltonian},
one obtains
\begin{eqnarray}
\label{eq::Ham2}
H\!\!&=&\!\!H_{\text{kin}} + H_{\text{int}}^{\rm MF},
\\
H_{\text{kin}}\!\!&=&\!\!\!\sum_{\langle ij\rangle\sigma}
	t_{ij} \left(c_{i\sigma}^{\dag}c^{\phantom{\dag}}_{j\sigma} + H.c.\right)
-
\sum_{i\sigma}\mu'_ic_{i\sigma}^{\dag}c^{\phantom{\dag}}_{i\sigma},
\\
H_{\text{int}}^{\rm MF}\!\!&=&\!\!\sum_{i}\!\left[
	\Delta_{i}
	(c_{i\downarrow}^{\dag}c^{\phantom{\dag}}_{i\downarrow}
	\!-\!
	c_{i\uparrow}^{\dag}c^{\phantom{\dag}}_{i\uparrow})
-U\!\left(\frac{x_i(2+x_i)}{4}-\frac{\Delta_{i}^2}{U^2}\!\right)\!\right]\!,\nonumber\\
\end{eqnarray}
where $\mu'_i=\mu-Ux_i/2$
is the effective (position-dependent) chemical potential, which accounts for both
$\mu$ and ``Hartree" contribution
$Ux_i/2$.
Parameters
$x_i$
and
$\Delta_i$
are to be found self-consistently.

\section{Phase separated states}
\label{sec::Inhomogeneous}

At zero doping the system's ground state is the homogeneous commensurate AFM.
When electrons or holes are added, the homogeneous state may become
unstable. In this section we will investigate two specific scenarious of
this instability. The simplest one is the phase separation into
commensurate AFM and paramagnetic phases. Required calculations are
straightforward and can be carried out analytically. A more comprehensive
approach is to take the incommensurate AFM state into account.
Corresponding calculations require numerical tools, however, the energy
estimate is improved. We will see that the phase separation into
incommensurate and commensurate AFM phases is more energetically favorable
than the commensurate AFM/paramagnet separation.

\subsection{Commensurate antiferromagnetism}
\label{subsec::commens_AFM}

We start our exposition with the simplest scenario: we assume that the
doped system remains homogeneous
$x_i=x=const$,
and the order parameter is commensurate:
\begin{eqnarray}
\label{eq::AFM_space}
\Delta_i=(-1)^{i_x+i_y+i_z} \Delta ,
\end{eqnarray}
where
$\Delta = const$,
and integers
$i_x$,
$i_y$,
and
$i_z$
describe the position of the lattice site
$i=(i_x,\,i_y,\,i_z)$.
The calculations presented below will prove that such a state is unstable.

Taking into account
Eq.~(\ref{eq::AFM_space}),
we make use of the Fourier transform to derive
\begin{eqnarray}
H &=&\!\sum_{\mathbf{k}\sigma}\left(\varepsilon_{\bf k} - \mu'\right)c_{{\bf k}\sigma}^{\dag}c^{\phantom{\dag}}_{{\bf k}\sigma}+\nonumber\\
&&+\!\sum_{\mathbf{k}}\Delta\left(c_{{\bf k}\uparrow}^{\dag}c^{\phantom{\dag}}_{{\bf k}+{\bf Q}_0\uparrow}-
c_{{\bf k}\downarrow}^{\dag}c^{\phantom{\dag}}_{{\bf k}+{\bf Q}_0\downarrow}\right)-\nonumber\\
&&-U{\cal N}\left(\frac{x(2+x)}{4}-\frac{\Delta^2}{U^2}\right),
\label{eq::Hamiltonian_fourier_nesting}	
\end{eqnarray}
where
$\mu'=\mu-Ux/2$,
the number of sites in the lattice is denoted by
${\cal N}$,
vector
${\bf k} = (k_x,k_y,k_z)$
is the quasi-momentum,
and
\begin{equation}
\label{eq::kin_en}
\varepsilon_{\bf k} = -2\left[t_x\cos(k_x)+t_y\cos(k_y)+t_z\cos(k_z)\right]
\end{equation}
is the kinetic energy. At half-filling
($x=0$, $\mu=\mu'\equiv0$)
the model's Fermi surface nests perfectly, with
${\bf Q}_0 = (\pi,\,\pi,\,\pi)$
as the nesting vector. Indeed, at the half-filling the Fermi surface is
defined by the equation
\begin{eqnarray}
\varepsilon_{\bf k} = 0,
\end{eqnarray}
which remains invariant under translation by
${\bf Q}_0$,
as guaranteed by the relation
\begin{eqnarray}
\label{eq::nesting}
\varepsilon_{{\bf k} + {\bf Q}_0} = -\varepsilon_{\bf k}.
\end{eqnarray}
While nesting is insensitive to the hopping anisotropy, it is destroyed by
both longer-range hopping and finite $\mu$, or, equivalently, finite
doping. It is the destruction of nesting by extra carriers that ultimately
destabilizes the homogeneous state postulated at the beginning of this
subsection.

To proceed with the solution, we introduce the four-component vector
$\psi^{\dag}_{\mathbf{k}}=(c^{\dag}_{\mathbf{k}\uparrow},\,c^{\dag}_{\mathbf{k}+\mathbf{Q}_0\uparrow},\,
c^{\dag}_{\mathbf{k}\downarrow},\,c^{\dag}_{\mathbf{k}+\mathbf{Q}_0\downarrow})$.
In terms of this vector, the Hamiltonian~\eqref{eq::Hamiltonian_fourier_nesting} can be written as
\begin{eqnarray}
H=\sum_{\mathbf{k}}\psi^{\dag}_{\mathbf{k}}\hat{H}_{\mathbf{k}}\psi^{\phantom{\dag}}_{\mathbf{k}}
-U{\cal N}\left(\frac{x(2+x)}{4}-\frac{\Delta^2}{U^2}\right),
\end{eqnarray}
where
$4\times4$
matrix
$\hat{H}_{\mathbf{k}}$
is
\begin{eqnarray}
\hat{H}_{\bf k}\!=\!
\begin{pmatrix}
\varepsilon_{\bf k}-\mu' & \Delta & 0 & 0\\
\Delta & \varepsilon_{{\bf k}+{\bf Q}_0} - \mu' & 0 & 0\\
0 & 0 & \varepsilon_{\bf k} - \mu' & -\Delta \\
0 & 0 & -\Delta & \varepsilon_{{\bf k}+{\bf Q}_0}-\mu'	
\end{pmatrix}.
\label{matrix::H2}
\end{eqnarray}
Using
Eq.~(\ref{eq::nesting}),
we express the eigenenergies as
\begin{eqnarray}
E_{\mathbf{k}}^{(1,2)}=-\mu'\mp\sqrt{\varepsilon_{\bf k}^2+\Delta^2}\,.
\end{eqnarray}
At zero temperature, the grand potential per site is
\begin{equation}
\label{eq::Omega}
\Omega=\sum_{s}\!\int\!\frac{d^3{\bf k}}{(2\pi)^3}E^{(s)}_{\mathbf{k}}\Theta(-E^{(s)}_{\mathbf{k}})-U\left(\frac{x(2+x)}{4}-\frac{\Delta^2}{U^2}\right)\!.	
\end{equation}
Minimizing $\Omega$ with respect to $\Delta$ we obtain the equation for the
order parameter. In the weak coupling limit, it is possible to solve this equation analytically (details are presented in Appendix~\ref{app::cAFM_formalism}). This gives the following equation relating the gap $\Delta$ and the chemical potential $\mu'$:
\begin{eqnarray}
\label{eq::13}
|\mu'| + \sqrt{\mu'^2 - \Delta^2} = \Delta_0\,,
\end{eqnarray}
where $\Delta_0$ is the gap at half-filling [see Eq.~\eqref{eq::Delta0} in Appendix~\ref{app::cAFM_formalism}]. Since a typical experiment is performed at fixed doping, not fixed chemical
potential, it is necessary to express the order parameter and the chemical
potential as functions of the doping level $x$.
The doping $x$ is given by the following relation:
\begin{eqnarray}
x=\sum _{s}\!\int\!\frac{d^3{\bf k}}{(2\pi)^3}\,\Theta(-E^{(s)}_{\mathbf{k}})-1\,.
\end{eqnarray}
Acting in the same manner as described in Appendix~\ref{app::cAFM_formalism}, we obtain in the weak coupling limit
\begin{eqnarray}
\label{eq::doping0}
x =2\rho_0\sign(\mu')\sqrt{\mu'^2-\Delta^2}\,,
\end{eqnarray}
where $\rho_0=\rho(0)$ is the density of states at zero energy [Fermi level at half-filling, see Eq.~\eqref{eq::rho} for definition of $\rho(E)$].
If we set
$\Delta = 0$
in
Eq.~\eqref{eq::doping0},
we recover a familiar expression
\begin{eqnarray}
\label{eq::mu_pm}
\mu = \frac{x}{2\rho_0},
\end{eqnarray}
which relates doping and chemical potential in the paramagnetic phase. In
Eq.~(\ref{eq::mu_pm})
the contribution
$Ux/2$
to the effective potential is omitted. The effects due to this term are
small in the weak coupling limit, as we will show below.

Using
equations~\eqref{eq::13}
and~\eqref{eq::doping0},
and neglecting
$Ux/2$
contribution to $\mu'$, we further
obtain~\cite{rice1970,tokatly1992}
\begin{eqnarray}
\label{eq::mu}
|\mu | &=& \Delta_0\left(1-\frac{|x|}{2\rho_0\Delta_0}\right),
\\
\label{eq::Delta}
\Delta&=&\Delta_0\sqrt{1-\frac{|x|}{\rho_0\Delta_0}}\,.
\end{eqnarray}
These two formulas describe homogeneous AFM state. We note that the
chemical potential is the decreasing function of the doping
\begin{eqnarray}
\frac{\partial\mu}{\partial x} = - \frac{1}{2\rho_0}<0\,.
\end{eqnarray}
It means that the compressibility is negative and homogeneous AFM state is
unstable, as announced above. This is the first example of the phase
separation. Observe that the small correction to
$\partial \mu/\partial x$
due to the omitted
$Ux/2$
contribution to $\mu'$ cannot restore the stability of the homogeneous
state as long as we consider the weak coupling limit.

\begin{figure}[t]
\centering{\includegraphics[width=0.99\columnwidth]{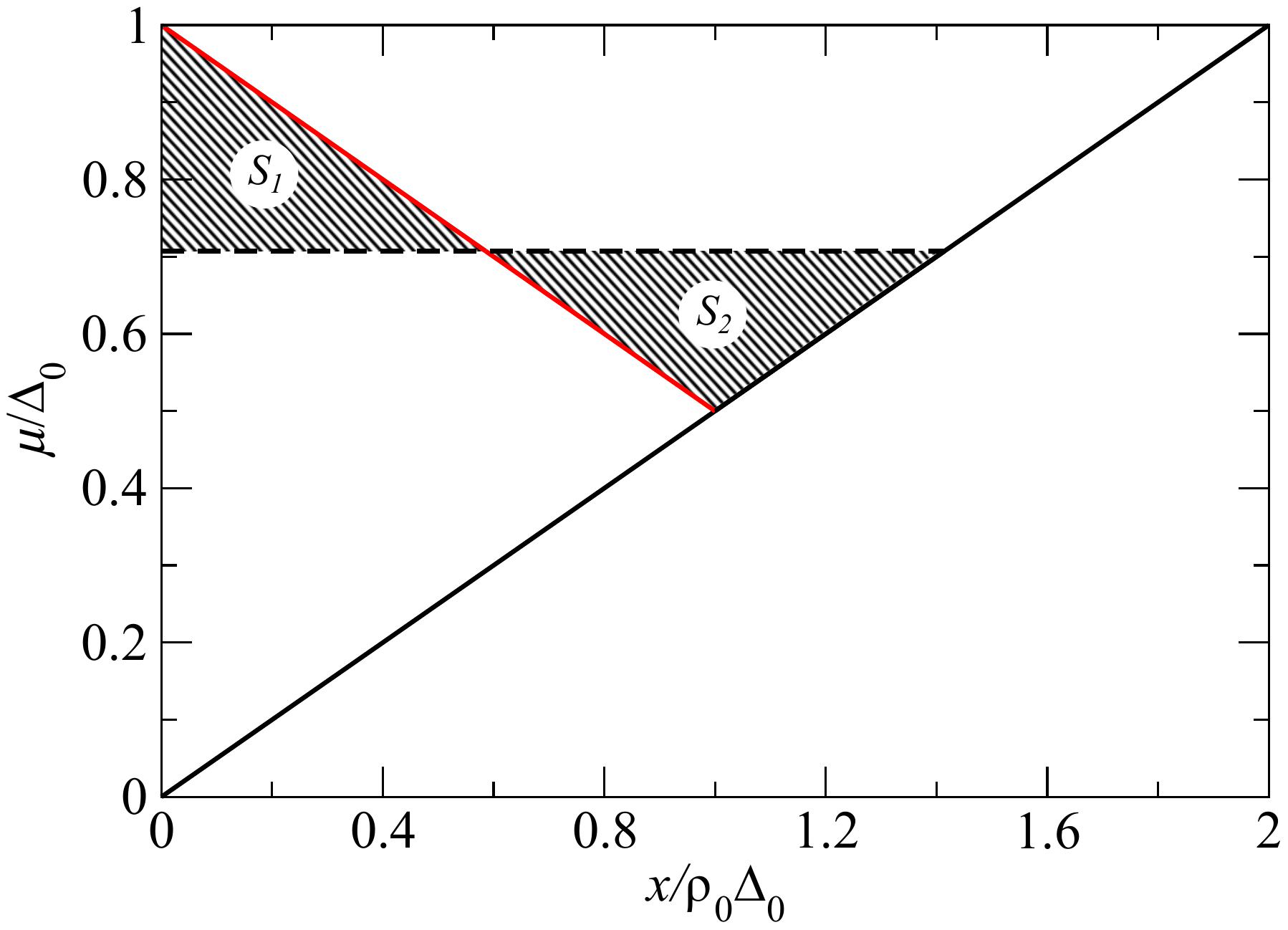}}
\caption{The doping dependence of the chemical potential for the
commensurate AFM and paramagnetic phases. Chemical potential $\mu$ and
doping $x$ are normalized by
$\Delta_0$
and
$\rho_0\Delta_0$,
correspondingly. From
$x=0$
to
$x=\rho_0\Delta_0/2$
the chemical potential in the commensurate AFM phase is shown as a straight
(red) line, see
Eq.~(\ref{eq::mu}).
Since this line has negative derivative, the doped commensurate AFM state
is unstable. For the paramagnetic phase,
$\mu = \mu (x)$
is shown as a straight (black) line, see
Eq.~(\ref{eq::mu_pm}).
To determine the chemical potential of the phase-separated state, we use
Maxwell construction: the horizontal dashed line is drawn to guarantee the
equality of the areas
$S_1=S_2$.
The chemical potential of the inhomogeneous state is
$\Delta_0/\sqrt{2}$.
As one can see from the Maxwell construction, the separation occurs into
the undoped AFM and the paramagnetic phases.
\label{fig::Maxwell_com}
}
\end{figure}

The structure of inhomogeneous phase can be established with the help of
Maxwell construction, see
Fig.~\ref{fig::Maxwell_com}.
It shows the chemical potential of the homogeneous commensurate AFM state
[decreasing line,
Eq.~(\ref{eq::mu})]
and paramagnetic state [increasing line,
Eq.~(\ref{eq::mu_pm})]
versus the doping level. The horizontal line in
Fig.~\ref{fig::Maxwell_com}
should be drawn so that areas
$S_1$
and
$S_2$
are equal. This line represents the phase-separated state. In thermodynamic
equilibrium (that is, neglecting metastabile states), the chemical
potential of the this state is
\begin{eqnarray}
\label{eq::mu_cAF}
\mu_{\rm cAF}=\frac{\Delta_0}{\sqrt{2}} \approx 0.707 \Delta_0,
\end{eqnarray}
where the subscript `cAF' stands for `commensurate antiferromagnet'. The
neglected
$Ux/2$
term introduces small correction (of the order of
$U \rho_0 \Delta_0$)
to the value of
$\mu_{\rm cAF}$.

The physical meaning of
$\mu_{\rm cAF}$
is the threshold value which must be exceeded by the chemical potential of
an external reservoir for doping to commence. The electrons injected into
the AFM parent state, however, do not spread over the whole lattice evenly.
Instead, as the Maxwell construction implies, the inhomogeneous state is
split into areas of the undoped AFM and paramagnetic areas (the latter
accumulate all the doping).

The main goal of the study is to determine which of the inhomogeneous
states is energetically favorable. At fixed doping, this can be decided by
comparison of the free energies
$F(x) = \Omega + \mu x$
of the competing phases. For evaluation of
$F(x)$,
the following expression is useful
\begin{eqnarray}
F(x)=F(0)+\int\limits_0^x\mu(x)dx\,,
\label{eq::Free_en}
\end{eqnarray}
where
$F(0)$
is the free energy of undoped AFM insulator. Since the chemical potential
is doping-independent in the phase-separated state, we derive
\begin{eqnarray}
\label{eq::F_cAF}
F_{\rm cAF}(x) = F(0) + \mu_{\rm cAF}x \,.
\end{eqnarray}
This expression is valid for sufficiently low doping, as long as the system
remains on the horizontal line in
Fig.~\ref{fig::Maxwell_com}.
In the following sections,
$F_{\rm cAF}$
will be compared with the free energies of other inhomogeneous states.

\subsection{Incommensurate antiferromagnetism}
\label{sec::Incommensurate}

We have seen in the previous section that at half-filling perfect nesting
is realized at
${\bf Q_0}=\left( \pi,\pi,\pi\right)$.
For finite doping the perfect nesting is impossible, but the quality of
nesting may be improved if we consider incommensurate AFM, whose nesting
vector is
\begin{equation}
{\bf Q}=\left( \pi,\,\pi,\,\pi\right)+{\bf q} = {\bf Q}_0 + {\bf q}\,.
\end{equation}
In this expression,
${\bf q}$
is (small) incommensurability vector. Non-zero
$|{\bf q}|$
means that antiferromagnetic order parameter takes the form
\begin{equation}
\Delta_i =\Delta(-1)^{i_x+i_y+i_z}e^{i{\bf q}{\bf {r}}_i},
\end{equation}
where
$\mathbf{r}_i=(i_x,\,i_y,\,i_z)$.
The interaction part of the mean-field Hamiltonian in
$k$-space
becomes
\begin{eqnarray}
H_{\text{int}}^{\rm MF}&=&\!\sum_{\mathbf{k}}\Delta\left(c_{{\bf k}\uparrow}^{\dag}c^{\phantom{\dag}}_{{\bf k}+{\bf Q}\uparrow}
-c_{{\bf k}\downarrow}^{\dag}c^{\phantom{\dag}}_{{\bf k}+{\bf Q}\downarrow}\right)-\nonumber\\
&&-U{\cal N}\left(\frac{x(2+x)}{4}-\frac{\Delta^2}{U^2}\right),
\label{eq::Hamiltonian_fourier_nesting_incom}	
\end{eqnarray}
Taking into account the equation
$\varepsilon_{\bf k+Q} = -\varepsilon_{\bf k+q}$,
we can write the equations for eigenenergies
\begin{equation}
E^{(1,2)}_{\mathbf{k}} = -\mu'+\frac{\varepsilon_{\bf k}-\varepsilon_{\bf k+q}}{2} \mp \sqrt{\left(\frac{\varepsilon_{\bf k}+\varepsilon_{\bf k+q}}{2}\right)^2+\Delta^2}\,.
\label{eq::eig_en_delta_q}
\end{equation}

\begin{figure}[t]
\centering{\includegraphics[width=0.99\columnwidth]{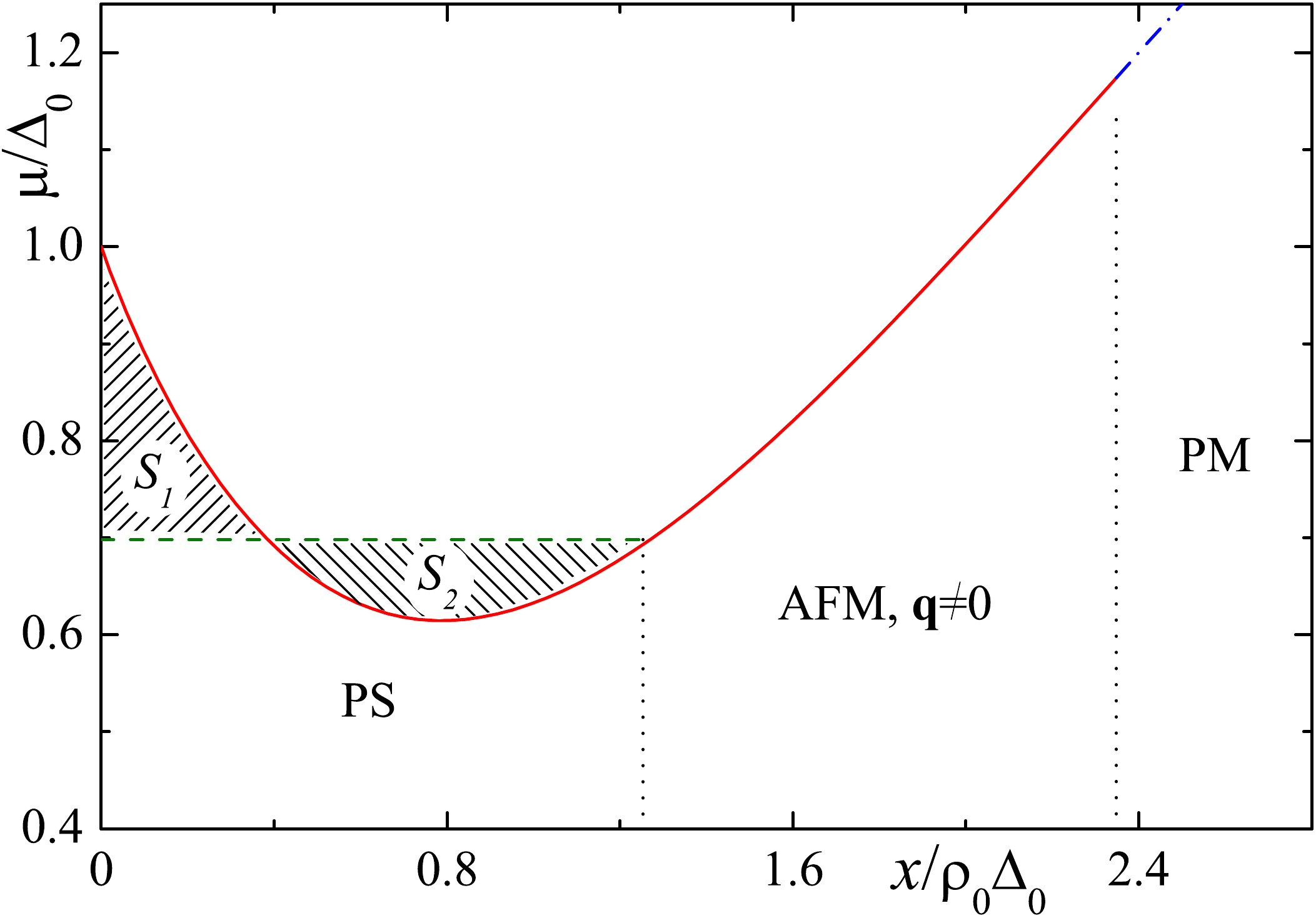}}
\caption{The doping dependence of the chemical potential of the
incommensurate AFM state (solid red curve) and the paramagnetic state (blue
dot-dashed curve). The Maxwell's construction requires that the hatched
areas
$S_{1,2}$
are equal to each other. Horizontal dashed (green) line corresponds to the
chemical potential of the phase-separated state. It equals to
$\mu_{\rm iAF}/\Delta_0\cong0.698$.
Model parameters are
$t_x=t_y=1$,
$t_z=0.7$,
incommensurability vector
$\mathbf{q}$
is parallel to the $z$ axis.
\label{fig::Maxwell_incomm}
}
\end{figure}

Grand potential of the system per one site is given by
Eq.~\eqref{eq::Omega}
with eigenenergies from
Eq.~\eqref{eq::eig_en_delta_q}.
The equations for the order parameter $\Delta$, nesting vector
$\mathbf{q}$, and the chemical potential are
\begin{equation}
\label{eq::min_Omega}
\frac{\partial\Omega}{\partial\Delta}=0\,,\;\;\frac{\partial\Omega}{\partial\mathbf{q}}=0\,,\;\; -\frac{\partial\Omega}{\partial\mu'}=1+x\,.
\end{equation}
These equations are soloved in the limit of small
$|{\bf q}|$.
The details of the calculations can be found in
Refs.~\onlinecite{rice1970,ourPRB_phasepAFM2017,phasep_pnics2013}
(for reader's convenience, they are outlined in
Appendix~\ref{app::AFM_formalism}
as well).

The resultant dependence
$\mu = \mu (x)$
calculated for $\mathbf{q}$ parallel to the $z$ axis is plotted in
Fig.~\ref{fig::Maxwell_incomm} [hopping amplitudes corresponding to this figure are $(t_x,\,t_y,\,t_z)=(1.0,\,1.0,\,0.7)$].
Here, as in the previous section, the small correction
$Ux/2$
was neglected.  We see non-monotonous behavior of
$\mu(x)$,
indicating the instability of the homogeneous state toward the phase
separation, analogous to what
Fig.~\ref{fig::Maxwell_com}
has shown. This time, however, the separated phases are (undoped)
commensurate and incommensurate AFM states, as one can prove using the
Maxwell construction. The chemical potential of the inhomogeneous state is
\begin{eqnarray}
\label{eq::mu_iAF}
\mu_{\rm iAF} \approx0.698\Delta_0,
\end{eqnarray}
where the subscript `iAF' stands for `incommensurate antiferromagnet'.
When
$\mathbf{q}$
is parallel to the $y$ axis, the dependence
$\mu=\mu(x)$
is very similar to that shown in
Fig.~\ref{fig::Maxwell_incomm},
except that the transition to the paramagnetic state in this case occurs at
smaller doping.

Similar to
$F_{\rm cAF}$,
Eq.~(\ref{eq::F_cAF}),
the free energy
$F_{\rm iAF}$
for the inhomogeneous phase represented by the horizontal line in
Fig.~\ref{fig::Maxwell_incomm}
is
\begin{eqnarray}
\label{eq::F_iAF}
F_{\rm iAF} (x) = F(0) + \mu_{\rm iAF} x.
\end{eqnarray}
It is easy to see that
\begin{eqnarray}
\label{eq::iaf_vs_caf}
F_{\rm iAF} (x) < F_{\rm cAF}(x)
\quad
\Leftrightarrow
\quad
\mu_{\rm iAF} < \mu_{\rm cAF}.
\end{eqnarray}
Thus, the phase separation into the commensurate and incommensurate AFM
phases is more favorable than the separation into the commensurate AFM
state and the paramagnetic state.

It is interesting to note that
Eq.~(\ref{eq::iaf_vs_caf})
reduces the comparison of the free energies to the comparison of the
threshold chemical potentials
$\mu_{\rm cAF}$
and
$\mu_{\rm iAF}$.
Since these quantities are very close to each other
($\mu_{\rm iAF} \cong0.698\Delta_0$
versus
$\mu_{\rm cAF} \cong0.707\Delta_0$),
the energy difference between these two inhomogeneous states is very small
for all relevant values of $x$.

\section{A state with domain walls}
\label{sec::Domain_wall}

\subsection{General considerations}

Yet another type of inhomogeneous phase competing to become the true
ground state is the phase with domain walls. In the previous section we
have seen that, to decide which phase-separated state is more energetically
favorable, the threshold chemical potentials have to be compared. In this
section, we will calculate
$\mu_{\rm dw}$,
the threshold chemical potential for the phase with domain walls.

When the system's chemical potential is close to the threshold value, the
doping concentration is low (this is a direct consequence of the threshold
chemical potential definition). A phase with domain walls in such a regime
is characterized by large inter-wall separation and negligible interaction
between the domain walls. Thus,
$\mu_{\rm dw}$
is determined by the properties of a single domain wall.

Preparing a study of a single domain wall properties, several
considerations must be taken into account. An important characteristics of
a domain wall is its orientation relative to lattice axes. The vector
normal to the domain wall plane may be parallel to one of the
crystallographic axes, or it may point in an arbitrary
direction~\cite{kato1990soliton}.
All these orientations cannot be investigated in complete generality, and
the study scope must be restricted. Numerical calculations for the
arbitrary orientations of the domain walls are computationally costly. We
expect that, in agreement with previous
publications~\cite{zheng2017stripe},
the domain walls whose normal vectors are parallel to one of the axis are
the most stable.

We study two types of domain walls: bond-centered and site-centered. They
can be schematically depicted with the help of the following
one-dimensional cartoons
\begin{eqnarray}
\nonumber
&&\uparrow \downarrow \uparrow
{ \downarrow \downarrow} \uparrow \downarrow \uparrow
\quad\text{bond-centered domain wall}, \\ \nonumber
&&\uparrow \downarrow \uparrow
{\downarrow {\rm o} \uparrow} \downarrow \uparrow \quad \text{site-centered
	domain wall}.
\end{eqnarray}
The arrows here represent the direction of the on-site spin magnetization,
the symbol `o' corresponds to a site with vanishing magnetization. As the
name implies, in the middle of the bond-centered wall, there is a bond
connecting two sites with identical magnetizations. A site with no net
magnetization is in the middle of the site-centered wall. Despite obvious
differences in real-space structures, our numerical simulations show that
the energies of bond-centered and site-centered configurations are very
close to each other.

\subsection{Mean-field description of a domain wall}

Let us now outline the mean-field formalism we employ to study a single
domain wall. For definiteness, we assume the domain wall is perpendicular
to the $x$-axis. For such an orientation, the translation invariance in $y$
and $z$ directions is preserved, while it is explicitly broken in
$x$-direction: the density of electrons and the order parameter are
\begin{eqnarray}	
&&\langle n_{i\uparrow}\rangle+\langle n_{i\downarrow}\rangle = n_{i_x},\,
\nonumber
\\
&&\Delta_i=\Delta_{i_x}(-1)^{i_y + i_z}\,.
\end{eqnarray}
Therefore, it is convenient to switch to the mixed representation: in the
$x$-direction, we continue using real space co-ordinate
$i_x$,
while in the transverse ($y$ and $z$) directions the 2D quasimomentum
$\mathbf{p}=(p_y,\,p_z)$
is introduced. We also define the partial Fourier transform of
$c^{\dag}_{i\sigma}$
as follows
\begin{eqnarray}
c^{\dag}_{i_x\mathbf{p}\sigma}
=
\frac{1}{\sqrt{{\cal N}_y {\cal N}_z}}
\sum_{i_y,i_z} c^{\dag}_{i\sigma} e^{i (p_y i_y + p_z i_z)},
\end{eqnarray}
where
${\cal N}_{\alpha}$
is the number of unit cells along axis
$\alpha = x, y, z$.

The mean-field Hamiltonian in the mixed representation reads
\begin{eqnarray}
H&=&H_{\text{kin}}+H_{\text{int}}^{\rm MF}\,,
\label{eq::Hamiltonian_fourier_full}
\\	
H_{\text{kin}}
&=&\!
\sum_{i_x\mathbf{p}\sigma}
	t_x\left(
		c^{\dag}_{i_x\mathbf{p}\sigma}
		c^{\phantom{\dag}}_{i_x+1\mathbf{p}\sigma}
		+
		H.c.
	\right)
+
\nonumber
\\
&&+\!
\sum_{i_x\mathbf{p}\sigma}
	\left(
		\varepsilon^\perp_{\mathbf{p}}-\mu'_{i_x}
	\right)
	c^{\dag}_{i_x\mathbf{p}\sigma}
	c^{\phantom{\dag}}_{i_x\mathbf{p}\sigma}\,,
\\	
H_{\text{int}}^{\rm MF}
&=&-\!
\sum_{i_x\mathbf{p}}
	\Delta_{i_x}\left(
		c^{\dag}_{i_x\mathbf{p}\uparrow}
		c^{\phantom{\dag}}_{i_x\mathbf{p}+\mathbf{P}_0\uparrow}
		-
		c^{\dag}_{i_x\mathbf{p}\downarrow}
		c^{\phantom{\dag}}_{i_x\mathbf{p}+\mathbf{P}_0\downarrow}
	\right)
\nonumber
\\
&&
-U\frac{\cal N}{{\cal N}_x}
\sum_{i_x} \left( \frac{n_{i_x}^2-1}{4}-\frac{\Delta_{i_x}^2}{U^2}\right),
\label{eq::Hamiltonian_fourier}
\end{eqnarray}
where
$\mu'_{i_x}=\mu-U(n_{i_x}-1)/2$.
We also use the notations
$\varepsilon^\perp_{\bf p} = -2t_{y}\cos(p_y)-2t_{z}\cos(p_z)$
and
$\mathbf{P}_0=(\pi,\,\pi)$.

If we introduce the
$2{\cal N}_x$-component vectors
\begin{eqnarray}
\psi^{\dag}_{\mathbf{p}\sigma}
=
\left(
	c^{\dag}_{1\mathbf{p}\sigma}\,,\dots,\,
	c^{\dag}_{{\cal N}_x\mathbf{p}\sigma}\,,
	c^{\dag}_{1\mathbf{p}+\mathbf{P}_0\sigma}\,,\dots,\,
	c^{\dag}_{{\cal N}_x\mathbf{p}+\mathbf{P}_0\sigma}
\right)\!,
\quad
\end{eqnarray}
the
Hamiltonian~\eqref{eq::Hamiltonian_fourier_full}
can be expressed as
$H=\sum_{\mathbf{p}\sigma}\psi^{\dag}_{\mathbf{p}\sigma}\hat{H}_{\mathbf{p}\sigma}\psi^{\phantom{\dag}}_{\mathbf{p}\sigma}$,
where the matrices
$\hat{H}_{\mathbf{p}\sigma}$
can be written in the following block form
\begin{gather}
\hat{H}_{{\bf p}\uparrow}=
	\begin{pmatrix}
		\hat{H}_{0{\bf p}}&\hat{\Delta}\\
		\hat{\Delta}&\hat{H}_{0{\bf p}+{\bf P}_0}
	\end{pmatrix},\;
\hat{H}_{{\bf p}\downarrow}=
	\begin{pmatrix}
		\hat{H}_{0{\bf p}}&-\hat{\Delta}\\
		-\hat{\Delta}&\hat{H}_{0{\bf p}+{\bf P}_0}
	\end{pmatrix}.
\end{gather}
In the latter formulas, the matrices
$\hat{H}_{0{\bf p}}$
and
$\hat{\Delta}$
are
\begin{eqnarray}
\hat{H}_{0{\bf p}}\!&=&\!
\begin{pmatrix}
\varepsilon^{\perp}_{\bf p}\!-\!\mu'_{1}\!&	
				t&	0&	\cdots&	t \\
t& 	
	\varepsilon^{\perp}_{\bf p}\!-\!\mu'_2\!&	
					t&	\cdots&	0\\
0&	t& 	\varepsilon^{\perp}_{\bf p}\!-\!\mu'_3\!&	
						\cdots&	0\\
\vdots&	\vdots&	\vdots&	\ddots&	\vdots\\
t&	0&	0&	\cdots&	
			\varepsilon^{\perp}_{\bf p}\!-\!\mu'_{{\cal N}_x}
\end{pmatrix}\!,\quad
\label{eq::H0psigma}
\\
\hat{\Delta}\!&=&\!\diag\left(\Delta_1,\,...,\,\Delta_{{\cal N}_x}\right)\!.
\end{eqnarray}
Constructing the matrix
$\hat{H}_{\mathbf{p}\sigma}$
we use periodic boundary conditions in the $x$-direction.

\begin{figure}[t]
\centering{\includegraphics[width=0.99\columnwidth]{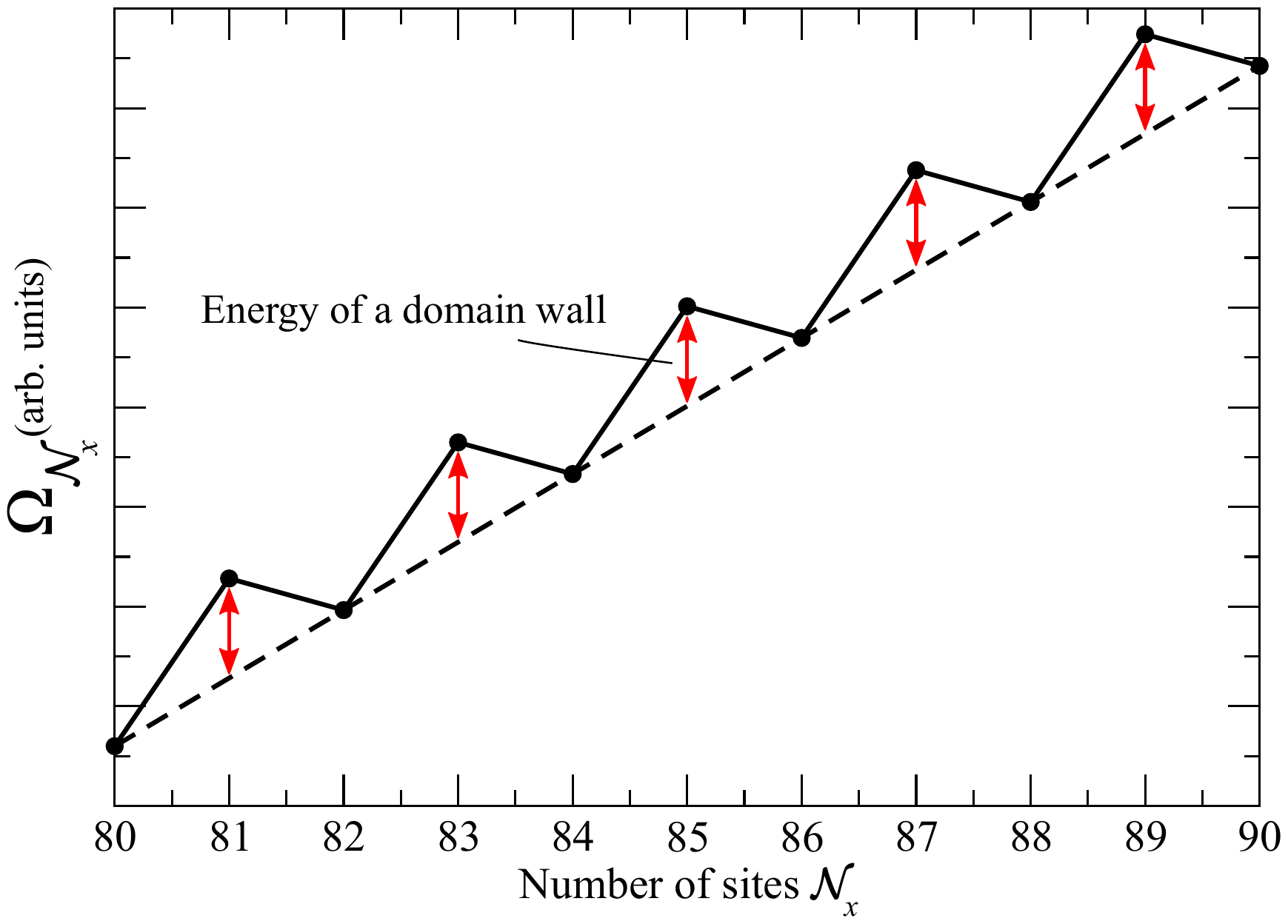}} 
\caption{Schematic illustration of the dependence of the grand
potential
$\Omega_{{\mathcal N}_x}$
on the number of sites in the $x$-direction. For even
${\cal N}_x$,
the dashed line represents proportionality
$\Omega^{\rm even}_{{\cal N}_x} \propto {{\cal N}_x}$,
see
Eq.~(\ref{eq::Omega_even}).
When
${\cal N}_x$
is odd,
$\Omega_{{\cal N}_x}$
is shifted by a constant value, see
Eq.~(\ref{eq::Omega_odd}).
This shift (shown by red arrows) is the energy of the domain wall
$E_{\rm dw}$.
\label{fig::dw_energy}
}
\end{figure}

For the system with
${\cal N}_x$
unit cells along
$x$-axis
the grand potential
$\Omega_{{\cal N}_x}$
(per unit area in
$y$-$z$~plane)
is
\begin{eqnarray}
\Omega_{{\cal N}_x}&=&\sum_{\sigma}\sum_{S=1}^{{\cal N}_x}
	\int\frac{d^2{\bf p}}{(2\pi)^2}
		E^{(S)}_{{\bf p}\sigma}\Theta(-E^{(S)}_{{\bf p}\sigma})+
\nonumber
\\
&& - U\sum_{i_x} \left(\frac{n_{i_x}^2-1}{4}-
	\frac{\Delta_{i_x}^2}{U^2}
\right),
\end{eqnarray}
where
$E^{(S)}_{{\bf p}\sigma}$
are the eigenenergies of the matrix
$\hat{H}_{\mathbf{p}\sigma}$.
To obtain
$\Omega_{{\cal N}_x}$,
the spatial dependencies of the order parameter
$\Delta_{i_x}$
and the number of electrons per site
$n_{i_x}$
minimizing
$\Omega_{{\cal N}_x}$
are found using a numerical recurrent procedure.

Once
$\Omega_{{\cal N}_x}$
is known, the energy of a single domain wall
$E_{\rm dw}$
can be calculated. To find
$E_{\rm dw}$,
it is necessary to consider systems with even and odd values of
${\cal N}_x$
(this number must be much larger than the width of the domain wall).
A system with even
${\cal N}_x$
is antiferromagnetically ordered and its grand potential (per unit area in
$y-z$~plane) is directly proportional to
${\cal N}_x$
\begin{eqnarray}
\label{eq::Omega_even}
\Omega_{{\cal N}_x}^{\rm even}=\Omega_0 {\cal N}_x\,,
\end{eqnarray}
where
$\Omega_0$
is the grand potential per site of the system with homogeneous AFM
ordering. A system with odd number of sites unavoidably contains a domain
wall. Therefore, the grand potential (per unit area) for the systems with
odd number of sites
$\Omega_{{\cal N}_x}^{\rm odd}$
can be expressed as
\begin{eqnarray}
\label{eq::Omega_odd}
\Omega_{{\cal N}_x}^{\rm odd} = \Omega_0 {\cal N}_x + E_{\rm dw}\,,
\end{eqnarray}
where
$E_{\rm dw}$
is the energy of the domain wall (per unit area in the transverse
directions). The
relations~(\ref{eq::Omega_even})
and~(\ref{eq::Omega_odd})
are illustrated in
Fig.~\ref{fig::dw_energy}.
They allow us to extract
$E_{\rm dw}$:
analysing
$\Omega^{\rm even}$
versus
${\cal N}_x$
dependence, we obtain
$\Omega_0$,
whose value is used to find
$E_{\rm dw}$
from the data for
$\Omega^{\rm odd}$.

\subsection{Numerical results}
\label{sec::Domain_wall_simulations}

\begin{figure}[t]
\centering{\includegraphics[width=0.99\columnwidth]{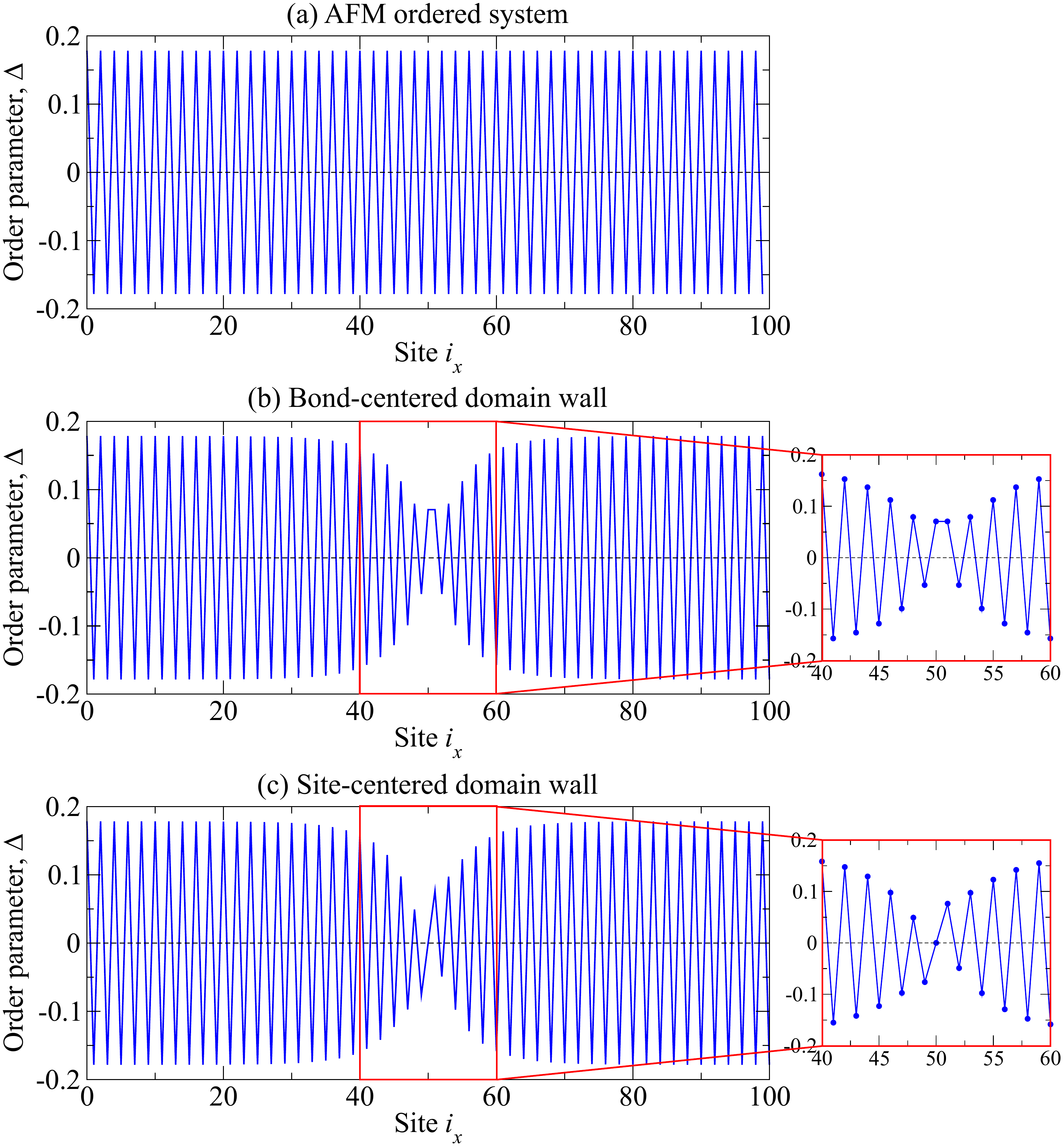}}
\caption{Spatial variation of the order parameter along $x$-axis,
for even and odd values of
${\cal N}_x$.
(a)~When the number of sites in $x$-direction
${\cal N}_x$
is even, the antiferromagnetic order parameter maintains the same absolute
value for all
$i_x$,
while the sign changes from one
$i_x$
to the next. The data in the panel is plotted for
${\cal N}_x = 100$.
(b, c)~When
${\cal N}_x$
is odd (specifically
${\cal N}_x = 101$
for both panels), the antiferromagnetic domain wall emerges. Panel~(b)
shows the bond-centered domain wall. In the center of such a structure, we
see two sites with identical values of the order parameters. The
site-centered domain wall is in panel~(c). This domain wall is centered on
a site with vanishing order parameter. The simulations are performed for
$U/W=0.17$,
$(t_x,\,t_y,\,t_z)=(1.0,\,1.0,\,0.7)$.
\label{fig::order_parameter}
}
\end{figure}

Numerically minimizing the grand potential
$\Omega_{{\cal N}_x}$
we determine various properties of the studied system.
Figure~\ref{fig::order_parameter}
demonstrates the spatial dependence of the order parameter for even and odd
${\cal N}_x$.
As we can see from
Fig.~\ref{fig::order_parameter}(a),
the system with even number of sites has the homogeneous antiferromagnetic
order:
$\Delta_{i_x}$
has a constant absolute value and opposite signs on any pair of nearest
sites. Naturally, the grand potential for such a state satisfies
Eq.~(\ref{eq::Omega_even}).

Due to the periodic boundary conditions, a system with odd number of sites
cannot maintain unfrustrated antiferromagnetic order, and a domain wall
appears. Because of the order parameter frustration,
$\Delta_{i_x}$
is suppressed inside the domain wall, see
Figs.~\ref{fig::order_parameter}(b,c).
In our simulations, we can stabilize both bond-centered and site-centered
domain walls.
Figure~\ref{fig::order_parameter}(b)
illustrates the order parameter structure for the bond-centered domain
wall. Such a configuration possesses spatial reflection symmetry with
respect to the center of the bond connecting the sites with minimum values
of the order parameter. Site-centered domain wall is shown in
Fig.~\ref{fig::order_parameter}(c).
Spatial inversion relative to central site of the domain wall (the site
with vanishing order parameter), accompanied by the spin flip
$\mathbf{S}\to-\mathbf{S}$,
preserves the site-centered configuration. Since bond-centered and
site-centered domain walls have different symmetries, they represent
mutually excluding classes of the mean-field solutions. Thus, they must be
discussed separately. However, our simulations show that their energies are
close to each other.

\begin{figure}[t]
\centering{\includegraphics[width=0.99\columnwidth]{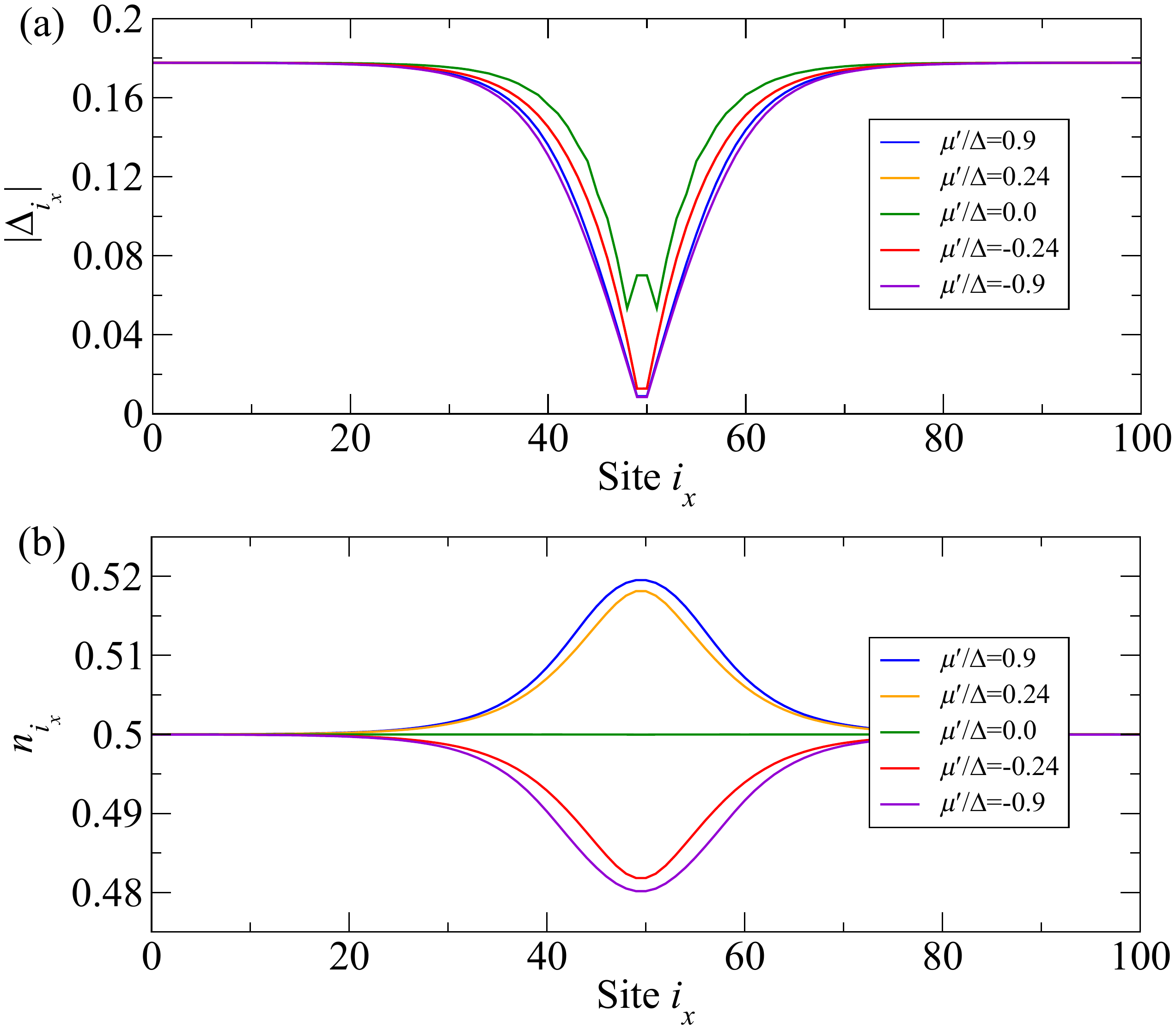}}
\caption{Spatial dependencies of (a)~the absolute values of the order
	parameter
$|\Delta_{i_x}|$,
and (b)~the electron density
$n_{i_x}$,
calculated for the bond-centered domain wall, at different values of the
chemical potential $\mu'$. Model parameters are:
$U/W=0.17$,
$(t_x,\,t_y,\,t_z)=(1.0,\,1.0,\,0.7)$,
and
${\cal N}_x=101$.
\label{fig::Chem_pot_delta_n}
}
\end{figure}

The domain wall properties are sensitive to the chemical potential. Indeed,
Fig.~\ref{fig::Chem_pot_delta_n}
demonstrates the spatial dependencies of the absolute value of the order
parameter and the electron density calculated for different values of
$\mu'=\mu-U(n-1)/2$
($n$ is the averaged number of electrons per site) for the system with
${\cal N}_x=101$,
$U/W=0.17$,
and
$(t_x,\,t_y,\,t_z)=(1.0,\,1.0,\,0.7)$.
One can see from this figure that even small deviation of the chemical
potential from zero value sharply changes the order parameter and the
electron density inside the domain wall. At higher values of
$|\mu'|$
the sensitivity of the order parameter and other quantities becomes less
dramatic.
Figure~\ref{fig::Chem_pot_delta_n}(a)
shows that the order parameter is even function of $\mu'$. Also, this
figure demonstrates that the domain wall becomes wider
when the chemical potential changes.

\begin{figure}[t]
\centering{\includegraphics[width=0.99\columnwidth]{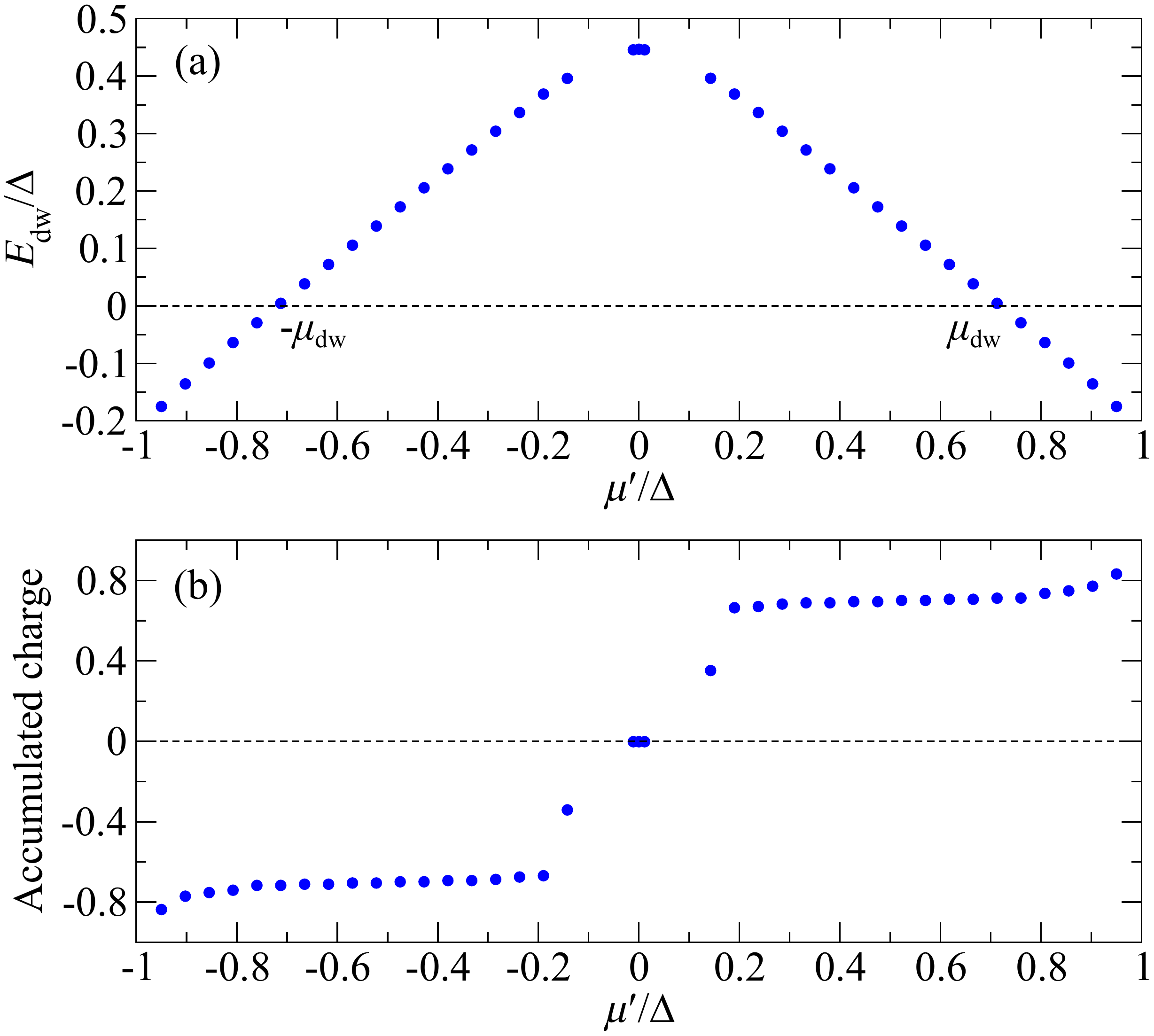}}
\caption{Chemical potential dependence of (a)~the bond-centered domain wall
energy
$E_{\rm dw}$
and (b)~the number of charge carriers in a domain wall. In panel~(a) the
critical chemical potential
$\mu_{\rm dw}$
is marked. It corresponds to the value of $\mu$ at which the domain wall
energy vanishes. Model parameters
are:
${\cal N}_x=101$,
$U/W=0.17$,
and
$(t_x,\,t_y,\,t_z)=(1.0,\,1.0,\,0.7)$.
\label{fig::Chem_pot_energy_doping}
}
\end{figure}

The accumulation of the injected charge carriers in the domain wall is
illustrated by
Fig.~\ref{fig::Chem_pot_delta_n}(b).
At half-filling
($\mu'=0$)
there is one electron per site. When the chemical potential changes, the
carriers pile up in the domain wall. For positive chemical potentials, the
carriers are electrons, and for the negative ones, they are holes. Finally,
Fig.~\ref{fig::Chem_pot_energy_doping}
presents the domain wall energy and the total charge accumulated inside
the domain walls (per unit area in $y$-$z$ plane) versus the
chemical potential in the system with
${\cal N}_x=101$,
$U/W=0.17$,
and
$(t_x,\,t_y,\,t_z)=(1.0,\,1.0,\,0.7)$.

What can be understood from the numerical data about the properties of a
single domain wall? We can see from
Fig.~\ref{fig::Chem_pot_energy_doping}
that the energy
$E_{\rm dw}$
is the even function of the chemical potential. Similarly,
Fig.~\ref{fig::Chem_pot_delta_n}(b)
and
Fig.~\ref{fig::Chem_pot_energy_doping}(b)
show that the accumulated charge is odd function of
$\mu'$.
These features are consequences of the charge-conjugation symmetry of our
model. This symmetry allows us to restrict our attention to positive value
of the chemical potential.

On general grounds, one expects that at zero doping and zero chemical
potential, the domain wall energy is positive, meaning that the state with
the domain walls is energetically unfavorable. However, as the chemical
potential grows, charges dope the domain walls, improving their stability.
Figures~\ref{fig::Chem_pot_energy_doping}(a,b)
clearly illustrate these tendencies. Most importantly, there is a specific
value of $\mu$ at which
$E_{\rm dw} = 0$.
When
$E_{\rm dw}$
vanishes, a state with no domain walls and a state with a domain wall are
degenerate. The corresponding value of $\mu$ is the critical chemical
potential
$\mu_{\rm dw}$
for the state with domain walls: if
$\mu > \mu_{\rm dw}$,
the domain wall energy becomes negative, and domain walls carrying finite
charge density enter the bulk of the system. As in the previous section,
\begin{eqnarray}
\label{eq::F_DW}
F_{\rm dw} \approx F(0) + \mu_{\rm dw} x
\end{eqnarray}
at low $x$. This equation remains applicable as long as the distance
between the domain walls is large, and the interaction between them may be
neglected. As the concentration grows, the repulsion between the walls sets
in, and
Eq.~(\ref{eq::F_DW})
progressively becomes less accurate. In the regime of validity of
Eqs.~(\ref{eq::F_cAF}),
(\ref{eq::F_iAF}),
and~(\ref{eq::F_DW})
the competition between the inhomogeneous states is decided by the lowest
critical chemical potential, as in
Eq.~(\ref{eq::iaf_vs_caf}).

\begin{figure}[t]
\centering{\includegraphics[width=0.99\columnwidth]{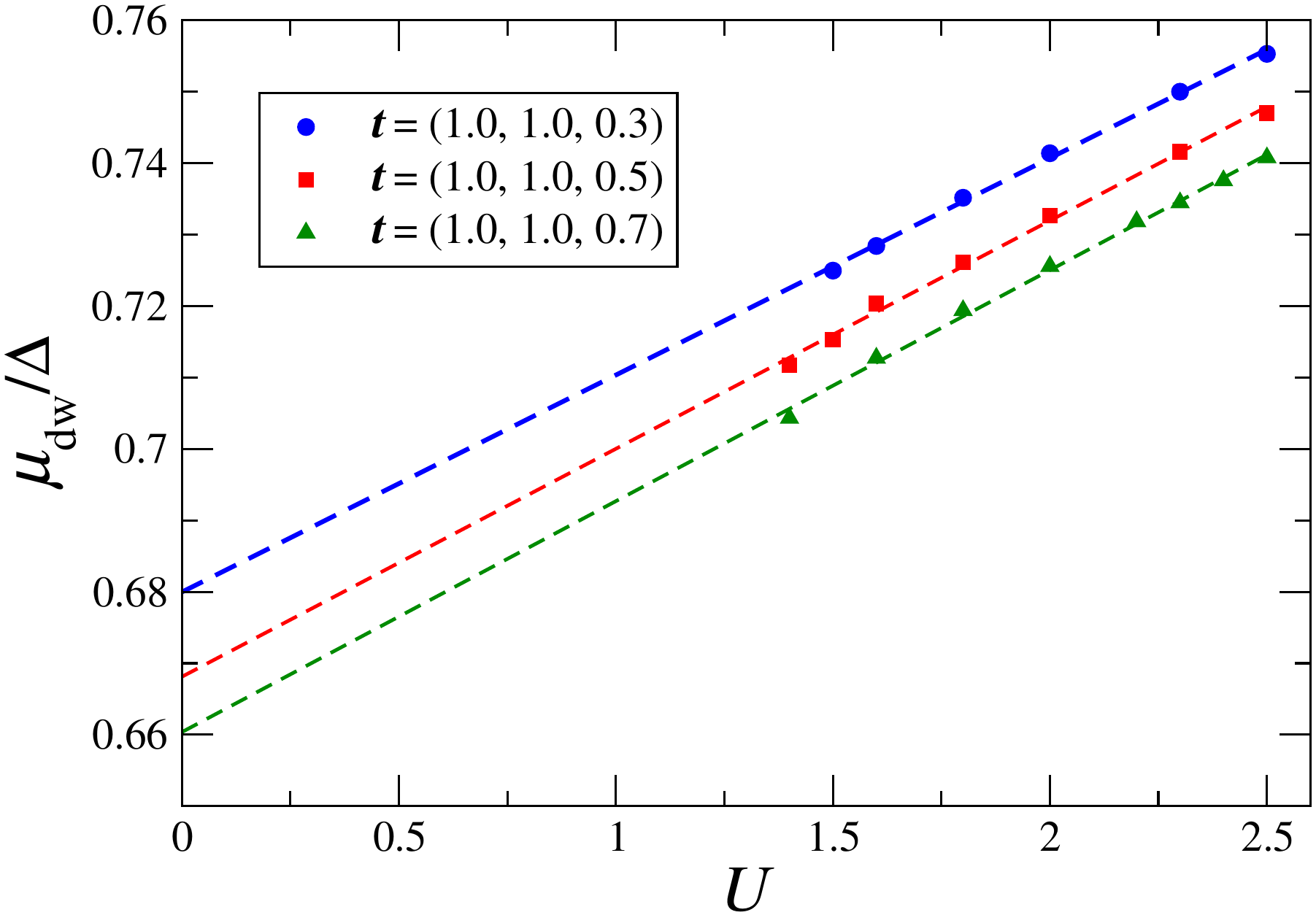}}
\caption{Critical chemical potential versus interaction constant $U$ for
three different anisotropies (see legend). Linear fits are shown as dashed
lines.
\label{fig::mu_c_vs_U}
}
\end{figure}

\begin{figure}[t]
\centering{\includegraphics[width=0.99\columnwidth]{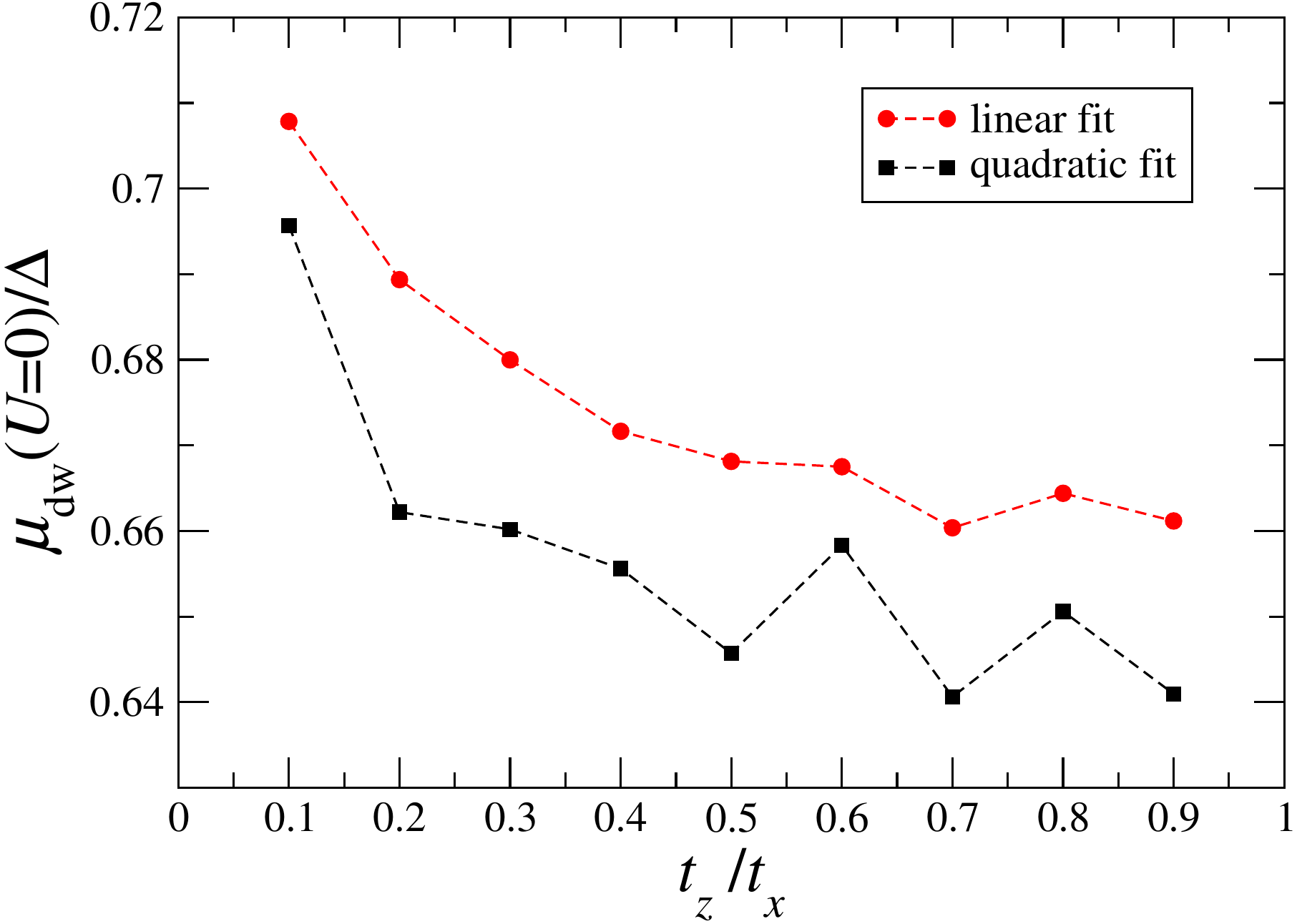}}
\caption{Critical chemical potential for the state with domain walls
$\mu_{\rm dw}$
versus the hopping anisotropy parameter
$t_z/t_x$,
in the limit of low $U$, for the anisotropy of the ``easy-plane"
($t_x = t_y > t_z$)
type. The data points obtained by linear fit (quadratic function fit) are
shown by red circles (black squares).
\label{fig::mu_c_vs_anisotrop}
}
\end{figure}

The critical potential calculated numerically is shown in
Fig.~\ref{fig::mu_c_vs_U}
for three different anisotropies of ``easy-plane"
($t_x = t_y > t_z$)
type. We see that the critical chemical potential monotonically increases
when the anisotropy increases. To perform consistent comparison of
$\mu_{\rm dw}$
with
$\mu_{\rm iAF}$
and
$\mu_{\rm cAF}$,
we need to find
$\mu_{\rm dw}$
in the low-$U$ limit, as we did above to obtain the
estimates~(\ref{eq::mu_cAF})
and~(\ref{eq::mu_iAF}).
Numerical calculations at very low $U$ quickly become impossible since the
width of the domain wall grows quickly as $U$ drops, and one has to
increase the system size stretching computational resources. To circumvent
this issue,
$\mu_{\rm dw}$
at
$U=0$
is evaluated extrapolating the available numerical data to zero value of
$U$. The data points can be adequately fitted by linear functions, see
Fig.~\ref{fig::mu_c_vs_U}.
The resultant low-$U$ values of
$\mu_{\rm dw}$
are shown in
Fig.~\ref{fig::mu_c_vs_anisotrop}.
Alternatively, the same data can be approximated by a quadratic function.
This produces similar values of
$\mu_{\rm dw}$,
which are also plotted in
Fig.~\ref{fig::mu_c_vs_anisotrop}.

The critical values of the chemical potential for different phases are
ordered according to
\begin{eqnarray}
\mu_{\rm dw} < \mu_{\rm iAF} \lesssim \mu_{\rm cAF}.
\end{eqnarray}
Thus, for the model under study, the state with the domain wall has the
lowest energy, at least at low doping. As we announced from the very
beginning, however, the energy difference separating the most stable phase
and metastable ``contenders" is insignificant. Indeed,
\begin{eqnarray}
\label{eq::006}
\mu_{\rm iAF} - \mu_{\rm dw} \lesssim 0.06 \Delta_0
\end{eqnarray}
for all anisotropy parameters.


\section{Discussion and conclusions}
\label{sec::Discussion}

In this paper we discuss inhomogeneous phases of the anisotropic Hubbard
model. To avoid uncontrollable approximations, we limit ourselves to the
weak coupling regime. Several interesting materials families, such as the
Bechgaard salts and iron-based superconductors, are characterized by weak
coupling. Electron-electron interaction in the cuprates superconductors is
believed to be strong or intermediate, therefore, our results are not
immediately applicable to the cuprates.

It is known that the homogeneous antiferromagnetic state of the Hubbard
model at one electron per site loses its stability upon doping. This
feature is not unique to the Hubbard Hamiltonian. Other models with nesting
demonstrate similar instability. In the context of superconductivity,
related phenomenon exists in the form of inhomogeneous
Fulde-Ferrel-Larkin-Ovchinnikov states. Thus, destruction of the electronic
liquid homogeneity is not limited to systems modelled by the Hubbard
Hamiltonian with repulsion, but rather is of interest in many situations.

We discuss three specific inhomogeneous phases (two types of phase
separated states and the state with domain walls) at zero temperature. It
is argued that, at low doping, the free energies of these states can be
characterized by a single parameter, critical chemical potential. The
latter concept has simple physical meaning: if the chemical potential of a
reservoir is lower than the critical chemical potential of a certain
inhomogeneous state, doping of our system through formation of this state
is impossible. It is clear from this definition that the phase with the
lowest critical chemical potential is the most stable at low doping.

The critical chemical potentials for all three phases were evaluated within
the mean field framework. Our calculations demonstrate that the state with
the domain wall (so-called ``soliton lattice") is the most stable. However,
all three values are close to each other, see
Eq.~(\ref{eq::006}).
This feature implies that purely theoretical prediction of the
inhomogeneous phase in the specific material is unreliable, as numerous
material factors lie outside of simple theoretical models. We expect that
the relative stability of various inhomogeneous states, competing to
become the true ground state, is affected by the lattice effects, band
structure details, Coulomb interaction screening, disorder pinning, and
other features. Reliable description of the inhomogeneous states
competition in specific materials appears to be impossible without
experimental input. The use of phenomenological models might be helpful as
well.

We already pointed out that our result is not directly applicable to the
cuprate superconductors, since the interaction constant in these materials
is not small. However, the theoretical issue addressed in this paper
remains relevant for the cuprates as well: it is quite possible that
several inhomogeneous states with almost identical energies compete against
each other obfuscating the interpretation of experimental data.

To conclude, we studied inhomogeneous states of the Hubbard model in
proximity to the half-filling. We demonstrated that the state with the
domain walls is the most stable at low doping. However, the energies of
metastable inhomogeneous states are found to be very close to the ground
state energy. We argue that the smallness of this energy separation can
introduce significant uncertainty into theoretical modeling of the
inhomogeneous states of many-body systems.

\begin{acknowledgments}
We acknowledge helpful discussions with B.V. Fine.
This work is partially supported by the JSPS-Russian Foundation for Basic
Research Project No.~19-52-50015,
and
RFBR Project RFBR no. 19-02-00421.
\end{acknowledgments}

\appendix

\section{Details of the solution to the gap equation for the commensurate AFM state}
\label{app::cAFM_formalism}

At half-filling ($\mu'=0$), minimization of the grand potential $\Omega$, Eq.~\eqref{eq::Omega}, with respect to $\Delta$ gives the following equation
\begin{equation}
\label{eq::gap0_eq}
\frac{\partial\Omega}{\partial\Delta}=
\Delta\left[\frac{2}{U}-\int\!\frac{d^3{\bf k}}{(2\pi)^3}\frac{1}{\sqrt{\Delta^2+\varepsilon_{\mathbf{k}}^2}}\right]=0\,.
\end{equation}
Let us introduce the density of states
\begin{equation}\label{eq::rho}
\rho(\varepsilon)=\int\!\frac{d^3{\bf k}}{(2\pi)^3}\,\delta(\varepsilon-\varepsilon_{\bf k})\,.
\end{equation}
In the limit of small
$\Delta\ll W$
studied here, we can rewrite the integral in the equation~\eqref{eq::gap0_eq} in the form
\begin{eqnarray}
&&\int\!\frac{d^3{\bf k}}{(2\pi)^3}\frac{1}{\sqrt{\Delta^2+\varepsilon_{\mathbf{k}}^2}}=\!\!\!\int\limits_{-W/2}^{W/2}\!\!\!d\varepsilon\,\frac{\rho(\varepsilon)}{\sqrt{\Delta^2+\varepsilon^2}}\cong\nonumber\\
&&\cong\!\!\!\int\limits_{-W/2}^{W/2}\!\!\!d\varepsilon\,\frac{\rho(\varepsilon)-\rho_0}{|\varepsilon|}+
\rho_0\!\!\!\int\limits_{-W/2}^{W/2}\!\!\!d\varepsilon\,\frac{1}{\sqrt{\Delta^2+\varepsilon^2}}\cong\nonumber\\
&&\cong\frac{2}{U_c}+2\rho_0\ln\frac{W}{\Delta}\,,\label{eq::IntAppr}
\end{eqnarray}
where $\rho_0=\rho(0)$ is the density of states at the Fermi level, while parameter $U_c$ is defined by the equation
\begin{equation}
\frac{2}{U_c}=\!\!\!\int\limits_{-W/2}^{W/2}\!\!\!d\varepsilon\,\frac{\rho(\varepsilon)-\rho_0}{|\varepsilon|}\,.
\end{equation}
Substituting Eq.~\eqref{eq::IntAppr} into Eq.~\eqref{eq::gap0_eq}, we obtain for the gap at half-filling:
\begin{equation}\label{eq::Delta0}
\Delta_0=W\exp\left[-\frac{1}{\rho_0}\left(\frac{1}{U}-\frac{1}{U_c}\right)\right].
\end{equation}
At finite doping, $\mu'$ deviates from zero, and the equation for the order
parameter can be expressed as (in the weak coupling limit)
\begin{equation}
2\ln\frac{\Delta_0}{\Delta}=\!\!\!\int\limits_{-W/2}^{W/2}\!\!\!d\varepsilon\,\frac{\Theta(|\mu'|-\sqrt{\Delta^2+\varepsilon^2})}{\sqrt{\Delta^2+\varepsilon^2}}\,.
\end{equation}
Evaluating the integral, one obtains Eq.~\eqref{eq::13} relating $\Delta$ and $\mu'$.

\section{Details of the mean-field formalism for the incommensurate AFM state}
\label{app::AFM_formalism}

In this Appendix, we will solve
Eqs.~(\ref{eq::min_Omega}).
We start with the observation that in the limit of small
$\Delta\ll W$,
nesting vector
${\bf q}$
is also small:
${\bf q}\sim\Delta/W\ll1$.
In this regime, one can write
\begin{equation}
\varepsilon_{\bf k+q}\cong\varepsilon_{\bf k}+{\bf q}\frac{\partial\varepsilon_{\bf k}}{\partial\mathbf{k}}\,.
\end{equation}
Calculating the derivatives of
$\Omega$
with respect to $\Delta$, $\mathbf{q}$, and $\mu'$, and using the smallness
of $\Delta$ and $\mathbf{q}$ in a manner similar to that described in the
Appendix~\ref{app::cAFM_formalism},
we obtain the system of equations:
\begin{eqnarray}
\ln\left(\frac{\Delta_0}{\Delta}\right) = \frac{1}{\rho_0}\!\!\int\limits_{-\eta_0}^{\eta_0}\!\!d\eta N(0,\eta)
\arccosh{\left( \frac{|\mu'-q\eta|}{\Delta}\right)},\quad
\label{eq::Delta_incom}
\\
q\kappa=\!\!\int\limits_{\eta_0}^{\eta_0}\!\!d\eta N(0,\eta)\eta\sqrt{(\mu'-q\eta)^2-\Delta^2}\sign(q\eta-\mu'),\quad
\label{eq::nesting_vec}
\\
x=2\!\!\int\limits_{\eta_0}^{\eta_0}\!\! d\eta N(0,\eta)\sqrt{(\mu'-q\eta)^2-\Delta^2}\sign(\mu'-q\eta),\quad
\label{eq::doping_incom}
\end{eqnarray}
where we introduce the joint density of states
\begin{eqnarray} 
N(\xi,\eta)=\int\frac{d^3\mathbf{k}}{(2\pi)^3}\,\delta\left(\xi-\varepsilon_{\bf k}\right)
\delta\left[\eta+{\hat n}_{\bf q}\frac{\partial\varepsilon_{\bf k}}{\partial\mathbf{k}}\right],
\\
\eta_0=\max_{\bf k}{\left({\hat n}_{\bf q}\frac{\partial\varepsilon_{\bf k}}{\partial\mathbf{k}}\right)},\;\;\kappa=\!\!\int\limits_{-\eta_0}^{\eta_0}\!\!d\eta N(0,\eta)\eta^2\,,
\end{eqnarray} 
and the unit vector
${\hat n}_{\bf q} = {\bf q}/|{\bf q}|$
is collinear with
$\mathbf{q}$.

Equations~\eqref{eq::Delta_incom},~\eqref{eq::nesting_vec},
and~\eqref{eq::doping_incom}
form a closed system of equations for self-consistent determination of
$\Delta(x)$,
$\mu(x)$,
and
$q(x)=|\mathbf{q}(x)|$
at the fixed direction of the nesting vector $\mathbf{q}$. We solve this
system of equations for
$(t_x,\,t_y,\,t_z)=(1,\,1,\,0.7)$,
and for two directions of
$\mathbf{q}$:
parallel to the $z$ axis and parallel to the $y$ axis. At relatively large
doping, the state with $\mathbf{q}$ parallel to the $z$ axis is
energetically more favorable, while at small doping the situation is
opposite. However, the difference in free energies between these two cases
turn out to be negligibly small. For the $z$ axis orientation, the data is
shown in
Fig.~\ref{fig::Maxwell_incomm}.


\end{document}